\documentclass[twocolumn,showpacs,prb]{revtex4}% Physical Review B
\usepackage{bm}
\usepackage{epsfig}
\newcommand{\nix}[1]{}

\begin{document}

\title{Electron and Hole Spin Splitting and Photogalvanic Effect in Quantum Wells}

\author{L.E.~Golub}

\email{golub@coherent.ioffe.ru}

\affiliation{A.F.~Ioffe Physico-Technical Institute, Russian
Academy of Sciences, 194021 St.~Petersburg, Russia}

%\date{\today}

\begin{abstract}
A theory of the circular photogalvanic effect caused by spin
splitting in quantum wells is developed. Direct interband
transitions between the hole and electron size-quantized subbands
are considered. It is shown that the photocurrent value and
direction depend strongly on the form of the spin-orbit
interaction. The currents induced by structure-, bulk-, and
interface-inversion asymmetry are investigated. The photocurrent
excitation spectra caused by spin splittings in both conduction
and valence bands are calculated.

\end{abstract}

\maketitle

\section{Introduction}
The spin properties of carriers have attracted much attention in
recent years due to rapidly developed spintronics dealing with the
manipulation of spin in electronic
devices.~\cite{spintronicbook02} The first idea was put forward by
Datta and Das, who proposed a spin field-effect
transistor.~\cite{spin_transistor} Its work is based on a change
of the Rashba field in semiconductor heterostructures, caused by
structure inversion asymmetry (SIA).~\cite{Bychkov84p78}

The promising materials for spintronics are III-V semiconductor
quantum wells~(QWs) whose spin properties are well documented and
can be controlled by advanced technology. However, in addition to
the Rashba spin-orbit interaction, other effective magnetic fields
act on carriers in zinc-blende heterostructures. This is, first,
the so-called Dresselhaus field caused by bulk inversion asymmetry
(BIA).~\cite{DK,Silva92,PikusPikus,Knap,PZ,SR-PRB} It is present
in QWs made of semiconductors without inversion center.

Second, there is the spin-orbit interaction produced by the
interface inversion asymmetry (IIA). It has been established that
the single-interface symmetry allows heavy-light hole mixing at
zero in-plane wave
vector.~\cite{Chang&Shulman,Aleiner,Kaminski,IKR,Krebs,Krebs1,Foreman,Cortez}
This mixing gives rise to spin splitting in both valence and
conduction bands at nonzero wave
vector.~\cite{Vervoort_PRB,Guettler,Vervoort_SST,R&K} The
structure of the corresponding linear in the wave vector
contribution to the electron effective Hamiltonian coincides with
that due to BIA.~\cite{R&K} In the following both contributions,
BIA and IIA, are considered together as a generalized BIA field.

Both SIA and BIA give rise to many spin-dependent phenomena in
QWs, such as an existence of beats in the Shubnikov-de~Haas
oscillations,~\cite{SdHspin} spin relaxation,~\cite{DK} splitting
in polarized Raman scattering spectra,~\cite{Jusserand} and
positive anomalous magnetoresistance.~\cite{WLspin} Spin
splittings and relaxation times have been extracted from these
experiments. However, in (001)-grown QWs, the SIA and BIA
spin-orbit interactions result in the same dependences of spin
splittings and spin relaxation times on the wave vector. Therefore
it is impossible to determine the nature for the spin splitting at
low wave vectors. Only in the simultaneous presence of both BIA
and SIA of comparable strengths, one can observe new effects, see
Ref.~\onlinecite{Review} and references therein. However the
latter situation is rare in real systems because it requires a
special structure design. Usually SIA and BIA cannot be
distinguished in experiments.

In this work, the other spin-dependent phenomenon is investigated
which is essentially different from the ones mentioned above. This
is the {\em Circular Photo-Galvanic Effect} which is a conversion
of photon angular momentum into a directed motion of charge
carriers. The result is an appearance of electric current under
absorption of a circularly-polarized light.~\cite{ELGE} The
photocurrent reverses its direction under inversion of the light
helicity. Microscopically, the circular photocurrent appears owing
to a coupling between orbital and spin degrees of freedom. In
semiconductors the coupling is a consequence of the spin-orbit
interaction. In two-dimensional systems, the circular
photogalvanic effect can be caused by both Rashba and Dresselhaus
effective magnetic fields. In this paper we show that SIA and BIA
result in experimentally distinguishable photocurrents.

The recently started activity on circular photogalvanics in QWs
attracted much attention.~\cite{Ganichev_last,Ganichev_Nature} The
photocurrents induced by both SIA and BIA have been investigated.
The circular photocurrent has been mostly studied under {\em
intra}band transitions induced by infrared or far-infrared
excitations. However it is important to extend studies on the
optical range where the effect is expected to be much stronger. In
this case, the photocurrent appears due to {\em inter}band
transitions.

In the present work, the theory of  the interband circular
photogalvanic effect in QWs is developed, and the photocurrent
spectra are calculated.

\section{General Theory}

In the absence of a spin-orbit interaction, two states with the
same wave vector $\bm{k}$ in the same size-quantized subband, $| 1
\rangle$ and $| 2 \rangle$, are degenerate. In a symmetrical QW,
the electron envelope functions of these states can be chosen in
the form
\begin{equation}
\label{el_w_f} | 1_e \rangle =  \exp{({\rm i} \: \bm{k} \cdot
\bm{\rho})} \: \varphi(z) \uparrow \:, | 2_e \rangle =
\exp{({\rm i} \: \bm{k} \cdot \bm{\rho})} \: \varphi(z) \downarrow \:,
\end{equation}
where $z$ is the growth direction, $\bm{\rho}$ is the electron
position in the plane of QW, and $\uparrow$ and $\downarrow$ are
the spin-up and spin-down functions. For odd (even) subbands of
size-quantization, $\varphi(-z) = \pm \varphi(z)$, respectively.

The two degenerate hole states with the wave vector ${\bm k}$ can
be chosen as symmetrical and antisymmetrical with respect to the
mirror reflection in the plane located in the middle of the QW.
The wave function for the symmetrical state has the
form~\cite{MPP}
\begin{eqnarray}
\label{h_w_f} && | 1_h \rangle = \exp{({\rm i} \: \bm{k} \cdot
\bm{\rho} \:)}\\ && \times N \left\{ - \left[ u_{3/2} + \sqrt{3}
W_+ \exp{(2 {\rm i} \phi_{\bm k})} u_{-1/2} \right] \chi_+(z)
\right. \nonumber
\\
&&+ \left.  {\rm i} \left[ \exp{(3 {\rm i} \phi_{\bm k})} u_{-3/2}
+ \sqrt{3} W_- \exp{({\rm i} \phi_{\bm k})} u_{1/2} \right]
\chi_-(z)\right\}, \nonumber
\end{eqnarray}
and the antisymmetrical wave function $| 2_h \rangle$ can be
obtained from $| 1_h \rangle$ by applying the operations of time
and space inversion. Here $u_{\pm 3/2}$, $u_{\pm 1/2}$ are the
Bloch functions at the top of the valence band, $\chi_+$ and
$\chi_-$ are, respectively, even and odd functions of coordinate
$z$,~\cite{zeta} and $\phi_{\bm k}$ is the angle between [100] and
${\bm k}$: $$\tan{\phi_{\bm k}} = k_y/k_x.$$

Spin-orbit interaction in QWs is described by the Hamiltonian
which has the following form in the basis of two states $| 1
\rangle$ and $| 2 \rangle$
\begin{equation}
\label{H} H({\bm k}) = {\bm \sigma} \cdot {\bm \Omega} ({\bm k})
\:.
\end{equation}
Here $\sigma_x$, $\sigma_y$ are the Pauli matrices acting on the
states $| 1 \rangle$ and $| 2 \rangle$. The effective
$\bm{k}$-dependent magnetic field ${\bm B}_{eff} \propto {\bm
\Omega} ({\bm k})$ is determined by the type of inversion
asymmetry.

In the presence of the spin-orbit interaction~(\ref{H}), the two
eigenstates with a given wave vector $\bm k$ are split by $\Delta
= 2 |{\bm \Omega} ({\bm k})|$. We denote these states by the index
$m = \pm$. Their envelope functions in size-quantized subbands can
be chosen in the form
\begin{equation}
\label{w_f_with_SOI} |m\rangle =  {m \over \sqrt{2}} \: \exp{(-
{\rm i} \Phi_{\bm k})} \: | 1 \rangle + {1 \over \sqrt{2}} \: | 2
\rangle \:,
\end{equation}
where the phase $\Phi_{\bm k}$ is given by $$\tan{\Phi_{\bm k}} =
\Omega_{y} / \Omega_{x}.$$

\begin{figure}
\epsfxsize=3in \epsfysize=3in \centering{\epsfbox{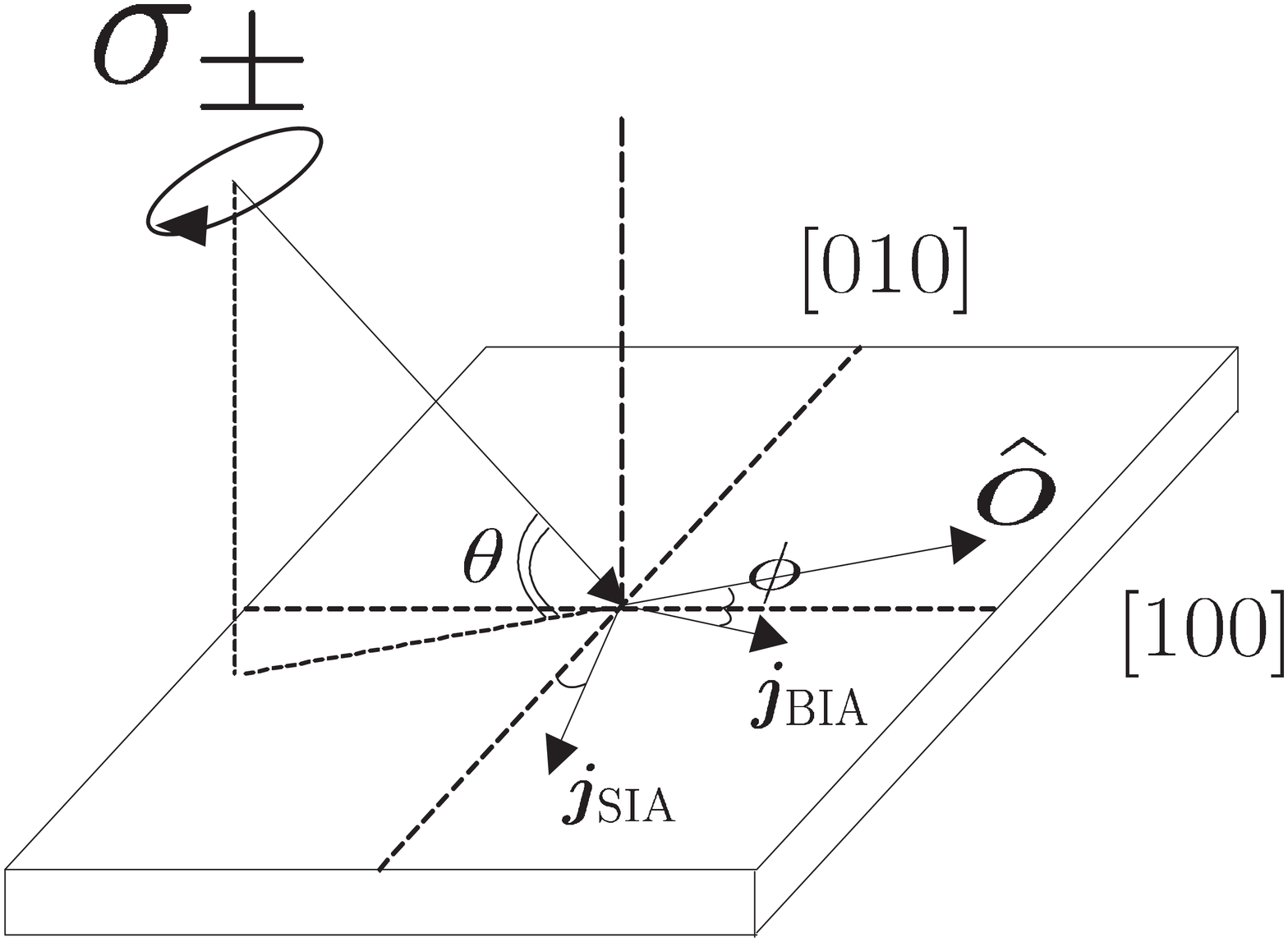}}
\caption{ \label{f0} SIA- and BIA-induced photocurrents appearing
under oblique incidence of circularly-polarized light.}
\end{figure}

Let us now consider optical excitation of a QW by
circularly-polarized light~(see Fig.~\ref{f0}).
Here we investigate the photocurrent arising due to asymmetry of
the carrier distribution in the ${\bm k}$-space at the moment of
creation. It is different from another photocurrent caused by
carrier momentum redistribution during the process of spin
relaxation. The latter, so-called spin-galvanic effect, was
considered in Refs.~\onlinecite{IL-GP_JETP}
and~\onlinecite{Ganichev_Nature}. These two photocurrents can be
separated in the time-resolved experiments: after switching-off
the light source, the former decays with the momentum relaxation
time while the latter disappears within the spin relaxation time.

The electric current is expressed in terms of the velocity
operators and spin density matrices for electrons and holes as
follows
\begin{equation}
\label{j} {\bm j} = e \sum_{\bm k} {\rm Tr} \left[ {\bm v}^{(e)}
({\bm k}) \: \rho^{(e)}({\bm k}) - {\bm v}^{(h)} ({\bm k}) \:
\rho^{(h)}({\bm k}) \right] \:,
\end{equation}
where $e$ is the electron charge.

Density-matrix equations taking into account both direct optical
transitions and elastic scattering give the following expressions
for linear in the light intensity values entering
into~Eq.(\ref{j})
\begin{eqnarray}
\label{rho} \rho^{(e,h)}_{nn'} = \pm {\pi \over \hbar} \tau_{e,h}
\sum_{\bar{n}} M_{n\bar{n}}M_{\bar{n}n'} &[\delta (E_n +
E_{\bar{n}} - \hbar \omega)& \\ + &\delta (E_{n'} + E_{\bar{n}} -
\hbar \omega)]& \nonumber \:.
\end{eqnarray}
Here $\tau_e$ and $\tau_h$ are the relaxation times of electrons
and holes (here we assume isotropic scattering), $\hbar \omega$ is
the photon energy, $M_{n\bar{n}}({\bm k})$ is the matrix element
of the direct optical transition between the states $n$ and
$\bar{n}$, and the energy dispersions $E_{e,h}({\bm k})$ are
reckoned inside the bands.

Spin splittings of electron or hole subbands both give rise to the
circular photogalvanic effect. The corresponding contributions to
the current are independent, therefore we consider the cases of
electron and hole spin splittings separately.

\section{Electron spin-splitting induced photocurrent}

Here we take into account spin splitting in the conduction band.
In QWs, the spin-orbit interaction Eq.~(\ref{H}) is described by
the linear in the wave vector Hamiltonian
\begin{equation}
\label{Omega_el} \Omega_i({\bm k}) = \beta_{il} k_l \:.
\end{equation}
Cubic in $\bm k$ corrections do not result in a circular
photogalvanic effect, therefore we do not consider them below.

Tensor $\beta$ is determined by the symmetry of the QW. As
mentioned in the Introduction, in structures with a zinc-blende
lattice, there is a contribution due to BIA and IIA known as the
Dresselhaus term. For (001)-grown QWs, it has two non-zero
components, namely
\begin{equation}
\label{BIA} \beta_{xx} = - \beta_{yy} \equiv \beta_{\rm BIA} \:,
\end{equation}
where $x \parallel [100]$ and $y \parallel [010]$.

SIA appears due to difference of the right and left barrier
heights of the QW, electric fields applied along $z$ direction,
etc. It leads to an additional contribution to the spin-orbit
Hamiltonian, the so-called Rashba term:
\begin{equation}
\label{SIA} \beta_{xy} = - \beta_{yx} \equiv \beta_{\rm SIA} \:.
\end{equation}

The velocity matrix elements calculated on the wave
functions~(\ref{w_f_with_SOI}) taking into account the spin-orbit
corrections~(\ref{H}), are given by
\begin{eqnarray}
\label{v} \left(v^{(e)}_i\right)_{mm'} &=& \left[ {\hbar k_i \over
m_e} + {m \over \hbar} (\beta_{xi} \cos{\Phi_{\bm k}} + \beta_{yi}
\sin{\Phi_{\bm k}}) \right] \delta_{mm'} \nonumber \\ &+& {{\rm i}
m \over \hbar} (\beta_{yi} \cos{\Phi_{\bm k}} - \beta_{xi}
\sin{\Phi_{\bm k}}) (1 - \delta_{mm'})\:, \nonumber
\\
v^{(h)}_i &=& {k_i \over k} v_h(k), \:\:\:\:  v_h(k) = {1 \over
\hbar} {d E_h(k) \over dk}\:,
\end{eqnarray}
where $m_e$ is the electron effective mass.

Calculations show that all odd Fourier harmonics of $\rho^{(e,h)}
({\bm k})$ entering into~(\ref{j}) are proportional to the
following part of the sum
\begin{eqnarray}
\label{rho_odd} && \sum_{l=|1_h \rangle, |2_h \rangle}
M_{ml}M_{lm'} \propto P_{circ} \left( e p A_0 \over m_0 c
\right)^2 \sin{\theta} \: (N Q_\pm)^2 \: m \nonumber \\
&& \times \left\{ \delta_{mm'} [W_\pm \cos{(\Phi_{\bm k} - 2
\phi_{\bm k} + \phi)} \right. + W_\pm^2 \cos{(\Phi_{\bm k} -
\phi)}] \nonumber \\
&& + {\rm i} \: (1 - \delta_{mm'} ) \\
&& \times \left. [W_\pm
\sin{(\Phi_{\bm k} - 2 \phi_{\bm k} + \phi)} + W_\pm^2
\sin{(\Phi_{\bm k} - \phi)}] \right\} \:. \nonumber
\end{eqnarray}
Here $P_{circ}$ is the circular polarization degree, $\theta$ and
$\phi$ are the spherical angles of the light polarization
vector~(see Fig.~\ref{f0}), $p$ is the interband momentum matrix
element, $A_0$ is the light wave amplitude, $m_0$ is the free
electron mass, and
\begin{equation}
\label{Q} Q_\pm = \int\limits_{-\infty}^{\infty} dz \: \varphi(z)
\: \chi_\pm(z) \:.
\end{equation}
The upper (lower) sign in Eq.~(\ref{rho_odd}) corresponds to
excitation into the odd (even) electron subbands.~\cite{zeta}

The characteristic spin splittings are usually very small,
therefore we consider a linear in $\beta$ regime. In this
approximation, SIA and BIA give independent contributions into the
photocurrent
\begin{equation}
\label{j_tot} {\bm j} (\omega) = {\bm j}_{\rm SIA} (\omega) + {\bm
j}_{\rm BIA} (\omega) \:,
\end{equation}
where ${\bm j}_{\rm SIA}$ and ${\bm j}_{\rm BIA}$ are linear in
$\beta_{\rm SIA}$ and $\beta_{\rm BIA}$, respectively. Assuming
the splitting $\Delta \to 0$ and calculating the reduced density
of states, we obtain from Eqs.~(\ref{j}) - (\ref{Q}) the
expressions for the interband circular photocurrents:
\begin{equation}
\label{j_fin} j_i (\omega) = - \beta_{li} \hat{o}_l \: P_{circ}
\left( e p A_0 \over m_0 \hbar c \right)^2 {e \over \hbar}  \:
G(k_\omega) \:.
\end{equation}
Here $i,l = x,y$, and $\hat{{\bm o}}$ is a projection of the unit
vector along the light propagation direction on the QW plane~(see
Fig.~\ref{f0}). The wave vector of the direct optical transition,
$k_\omega$, satisfies the energy conservation law
\begin{equation}
\label{k_omega}
E_e(k_\omega) + E_h(k_\omega) = \hbar \omega - E_g \:,
\end{equation}
where $E_e(k) = \hbar^2 k^2 / 2 m_e$ is the electron energy
without spin-orbit corrections, and $E_h(k)$ stands for the hole
dispersion, calculated in the spherical approximation.

We study the effects linear in $\beta$, therefore
Eq.~(\ref{j_fin}) is valid at $\hbar \omega - E_g^{QW} \gg \beta
k_\omega$, where $E_g^{QW} = E_g + E_{e1}(0) + E_{h1}(0)$ is the
fundamental energy gap corrected for the energies of
size-quantization. However for real systems, the theory is valid
even near the absorption edge.

The frequency dependence of the photocurrent is given by the
function $G(k)$
\begin{equation}
\label{G} G(k) = {k \over v(k)} {d \over dk} \left[ {F(k) u(k)
\over v(k)} \right] + {F(k) \over v(k)} \left[{u(k) \over v(k)} -
2 \right] \:,
\end{equation}
where
\begin{eqnarray}
\label{F&v&u} && F(k) = k [N(k) Q_\pm(k) W_\pm(k)]^2 \tau_e(k) \:, \\
&& v(k) = {\hbar k \over m_e} + v_h(k), \: u(k) = \left[ {\hbar k
\over m_e} + v_h(k) {\tau_h(k) \over \tau_e(k)} \right] \xi(k).
\nonumber
\end{eqnarray}
The first term in Eq.~(\ref{G}) appears because the direct
transitions to the upper (lower) spin branch take place at a wave
vector slightly smaller (larger) than $k_\omega$, and the second
term occurs because the two electron spin states with the same
$\bm k$ have different velocities.

The factor $\xi(k)$ depends on the form of a spin-orbit
interaction. It follows from Eqs.~(\ref{BIA}),~(\ref{SIA}) that,
for the BIA-induced spin-orbit interaction, $\Phi_{\bm k} = -
\phi_{\bm k}$, while, for SIA-dominance, $\Phi_{\bm k} = \phi_{\bm
k} - \pi/2$. Therefore one has
\begin{equation}
\label{xi} \xi_{\rm BIA}=1\:, \hspace{1cm} \xi_{\rm SIA}=1 -
1/W_\pm(k) \:.
\end{equation}
The difference appears because in the BIA-case the terms with
$W_\pm$ in Eq.~(\ref{rho_odd}) are the third harmonics of
$\phi_{\bm k}$ and, hence, do not contribute to the current. This
means that BIA and SIA create different current distributions of
optically-generated electrons.

This difference gives rise to non-equal frequency dependences of
the photocurrent. It is dramatic at the absorption edge, when
$\hbar \omega \geq E_g^{QW}$. For the ground hole subband $W_+(k)
\sim k^2$, and hence
\begin{equation}
\label{j_edge} j_{\rm BIA} \sim (\hbar \omega - E_g^{QW})^2 \:,
\hspace{1cm} j_{\rm SIA} \sim \hbar \omega - E_g^{QW} \:.
\end{equation}
This conclusion opens a possibility to distinguish experimentally
which kind of asymmetry, BIA or SIA, is dominant in the structure
under study. This could be done by studying the power, quadratic
or linear, in the dependence of the circular photocurrent on the
light frequency near the absorption edge. At higher photon
energies, the spectra are also different due to $k$-dependence of
the functions $W_\pm$ [see Eq.~(\ref{xi})].

\section{Hole spin-splitting induced photocurrent}

In this section we calculate the circular photocurrent which
arises due to the spin splitting in the valence size-quantized
subbands. The split hole states have wave functions
Eq.~(\ref{w_f_with_SOI}), the two degenerate states in the
conduction band are given by Eq.~(\ref{el_w_f}).

SIA and BIA result in the hole spin splitting as well as for
electrons. Below we consider both cases and derive the expressions
for circular photocurrents.

\subsection{SIA-induced circular photocurrent}

SIA is described by the hole
Hamiltonian~\cite{Winkler,Ganichev_SiGe} which has the following
form in the basis $u_{3/2}$, $u_{1/2}$, $u_{-1/2}$, $u_{-3/2}$

\begin{equation}\label{SIA_hole}
H({\bm k}) = - {\rm i} \alpha k \left(%
\begin{array}{cccc}
  0 & 0 & 0 & 0 \\
  0 & 0 & e^{- {\rm i} \phi_{\bm k}} & 0 \\
  0 & - e^{{\rm i} \phi_{\bm k}} & 0 & 0 \\
  0 & 0 & 0 & 0 \\
\end{array}%
\right).
\end{equation}
In a weak electric field applied along the growth direction, the
$k$-independent coefficient $\alpha$ is proportional to its
strength.

In the basis $|1_h\rangle$, $|2_h\rangle$ the Hamiltonian
Eq.~(\ref{SIA_hole}) has the form of Eq.~(\ref{H}) with
\begin{equation}\label{Omega_SIA_hole}
\Omega_x + {\rm i}\Omega_y = \langle 2_h | H |1_h\rangle = {\rm i}
A \exp{(3 {\rm i} \phi_{\bm k})},
\end{equation}
where
\begin{equation}\label{beta_SIA}
A = 3 \alpha k N^2 \left( W_+^2 \int\limits_{-\infty}^{\infty} dz
\: \chi_+^2(z) +  W_-^2 \int\limits_{-\infty}^{\infty} dz \:
\chi_-^2(z) \right).
\end{equation}
The splitting equals to $\Delta = 2 |A(k)|$, and the phase
$\Phi_{\bm k} = 3 \phi_{\bm k} + \pi/2$.

At $k \to 0$, for the $h1$ subband one has $\Delta \sim k^3$. For
the $l1$ subband, the Hamiltonian in the basis $\exp{(2 {\rm i}
\phi_{\bm k})} |2_h\rangle$, $\exp{(- 2 {\rm i} \phi_{\bm k})}
|1_h\rangle$ has the form
$$
    H({\bm k}) = \alpha (\sigma_x k_y - \sigma_y k_x),
$$
i.e. it is similar to the SIA-Hamiltonian for electrons,
Eqs.~(\ref{Omega_el}),~(\ref{SIA}). The splitting in this case is
linear in $k$. In real experiments, the hole wave vector may be
large, therefore we take into account the nonparabolicity effects
and allow explicit $k$-dependence $A(k)$ in
Eq.~(\ref{Omega_SIA_hole}).

Calculating the photocurrent, we arrive at the
expression~(\ref{j_fin}) with the $k$-dependent tensor $\beta$
given by
\begin{equation}\label{beta_SIA_holes}
\beta_{xy}(k) = - \beta_{yx}(k) = {d A \over dk}
\end{equation}
and with the following function $G(k)$ which determine the
photocurrent excitation spectrum
\begin{eqnarray}\label{G_SIA_holes}
    G(k) &=& {k \over \gamma(k) v(k)} {d \over d k} \left[ {F(k) u(k)
\over v(k)} \right] \\
&+& {F(k) \over v(k)} \left[{u(k) \over v(k)} - 2 + \left(1 + {1
\over W_\pm(k)} \right) \left(1 \pm {3 \over \gamma(k)}
\right)\right]. \nonumber
\end{eqnarray}
Here $F$, $u$ and $v$ are given by Eq.~(\ref{F&v&u}) with $\xi = 1
- 1/W_\pm$ and interchange $\tau_e \leftrightarrow \tau_h$. The
factor $\gamma$
\begin{equation}\label{gamma_SIA_holes}
    \gamma(k) = {d \ln{A} \over d \ln{k}}.
\end{equation}
At $k \to 0$, $\gamma = 3$ for the $h1$-subband and $\gamma = 1$
for the $l1$-subband. The edge behavior of the spectrum is
governed by the $h1 \to e1$ transitions and follows the quadratic
law
\begin{equation}
\label{j_edge_SIA_holes}
j_{\rm SIA} \sim (\hbar \omega - E_g^{QW})^2.
\end{equation}

\subsection{BIA- and IIA-induced circular photocurrent}

IIA is described by the interface term in the
Hamiltonian~\cite{Aleiner,Kaminski,IKR,Ganichev_SiGe,Phys_E}
\begin{equation}\label{H_IIA}
    H_{\rm IIA} = {\hbar^2 \over m_0 a_0 \sqrt{3}} \:
    t_{l-h}^{(i)} \{J_x,J_y\} \: \delta(z-z_i).
\end{equation}
Here $z_i$ are coordinates of the left ($i=L$) and the right
($i=R$) interfaces, $a_0$ is the lattice constant, and $J_{x,y}$
are the operators of angular momentum $3/2$. The heavy-light hole
mixing coefficients at two interfaces of A$_3$B$_5$ QWs
are related by
$$t_{l-h}^{(L)} = - t_{l-h}^{(R)}.$$

The BIA-contribution for holes in the bulk is described by the
Hamiltonian:~\cite{ELGE,{Mauritz97p10729}}
\begin{eqnarray}\label{H_holes_BIA_bulk}
    H_{\rm BIA} &=& {4 \over 3} c_1 \: \bm{K \cdot V} +
    c_3 \: [J_x K_x (K_y^2 - K_z^2) \\
    &+& J_y K_y (K_z^2 - K_x^2)+ J_z K_z (K_x^2 - K_y^2)]. \nonumber
\end{eqnarray}
Here $c_1$ and $c_3$ are the relativistically-small and
non-relativistic valence-band spin-orbit constants, respectively,
${\bm K} = ({\bm k}, -{\rm i} d/dz)$ is the three-dimensional wave
vector, and $V_x = \{J_x,(J_y^2 - J_z^2)\}$ etc. In
Eq.~(\ref{H_holes_BIA_bulk}) we neglected the cubic in $K$
relativistically-small terms.

In the basis $|1_h\rangle$, $|2_h\rangle$ both
Hamiltonians~(\ref{H_IIA}) and~(\ref{H_holes_BIA_bulk}) have the
form of Eq.~(\ref{H}) with
\begin{equation}\label{Omega_BIA_hole}
\Omega_x + {\rm i}\Omega_y = B_1 \exp{(5 {\rm i} \phi_{\bm k})} +
B_2 \exp{({\rm i} \phi_{\bm k})},
\end{equation}
where
\begin{eqnarray}\label{B}
&& B_{1,2} = \pm 3 N^2 \Biggl\{{2 \hbar^2 \over \sqrt{3} m_0 a_0}
W_\pm t_{l-h}^{(i)} \chi_+(z_i) \chi_-(z_i) \\
&+& c_3 \: k \left[W_\pm^2 \int\limits_{-\infty}^{\infty} dz \:
(d\chi_\pm / dz)^2 - W_\mp \int\limits_{-\infty}^{\infty} dz \:
(d\chi_\mp / dz)^2 \right] \nonumber \\
&+& \left[ c_3 {k^3 \over 4} (1- W_\pm) - c_1 \: k
\right] W_\pm^2 \int\limits_{-\infty}^{\infty} dz \: \chi_\pm^2(z) \nonumber \\
&+& \left[ c_3 {k^3 \over 4} (1- W_\mp) - c_1 \: k {1 + 2 W_\mp
\over 3} \right] \int\limits_{-\infty}^{\infty} dz \:
\chi_\mp^2(z) \Biggr\}. \nonumber
\end{eqnarray}
The splitting is given by
$$\Delta = 2 \sqrt{B_1^2 + B_2^2 + 2 B_1 B_2 \cos{4 \phi_{\bm k}}}
\:.$$ It is seen from Eq.~(\ref{B}) that IIA and BIA give additive
contributions to the spin-orbit interaction: the coefficients
$B_{1,2}$ have the terms proportional to $t_{l-h}^{(i)}$ and to
$c_{1,3}$. At $k \to 0$ for the $h1$ subband $B_1 \sim k^5$, $B_2 \sim k$.

Calculating the photocurrent, we arrive at the expression
\begin{eqnarray}
\label{j_BIA_holes} j_i (\omega) &=& - P_{circ} {e \over \hbar}
\left( e p A_0 \over m_0 \hbar c \right)^2  \\
&\times& \left[\beta_{li}^{(1)}(k_\omega)G_{1}(k_\omega) +
\beta_{li}^{(2)}(k_\omega)G_{2}(k_\omega) \right] \hat{o}_l .
\nonumber
\end{eqnarray}
The $k$-dependent tensors $\beta$ are given by
\begin{equation}\label{beta_BIA_holes}
\beta_{xx}^{(1,2)}(k) = - \beta_{yy}^{(1,2)}(k) = {d B_{1,2} \over
dk}
\end{equation}
and
\begin{eqnarray}\label{G_BIA_holes}
    G_{1,2}(k) &=& {k \over \gamma_{1,2}(k) v(k)} {d \over d k} \left[ {F_{1,2}(k) u(k)
\over v(k)} \right] \\
&+& {F_{1,2}(k) \over v(k)} \left[{u(k) \over v(k)} - 1 +
\nu_{1,2}(k)\right]. \nonumber
\end{eqnarray}
Here $u$ and $v$ are defined by Eq.~(\ref{F&v&u}) with $\xi = 1$
and interchange $\tau_e \leftrightarrow \tau_h$; $\gamma_{1,2} = d
\ln{B_{1,2}} / d \ln{k}$,
\begin{equation}\label{nu}
    \nu_1(k) = -5/\gamma_1(k), \hspace{1cm} \nu_2(k) =
    1/\gamma_2(k),
\end{equation}
\begin{eqnarray}\label{F_BIA_holes}
    F_1 &=& \pm k (N Q_\pm)^2 \tau_h W_\pm^{3/2 \pm 1/2}, \\
    F_2 &=& \pm k (N Q_\pm)^2 \tau_h W_\mp^{3/2 \mp 1/2}. \nonumber
\end{eqnarray}
The upper (lower) sign should be taken for transitions to the odd
(even) electron subbands.

At the absorption edge we have the linear law
\begin{equation}
\label{j_edge_BIA_holes}
j_{\rm BIA} \sim \hbar \omega - E_g^{QW}.
\end{equation}

The edge behavior of the photocurrent excitation spectra
Eqs.~(\ref{j_edge_SIA_holes}),~(\ref{j_edge_BIA_holes}) are
opposite to the electron case. The reason is in the complicated
valence band structure. For example, the parabolic
behavior~(\ref{j_edge_SIA_holes}) contrasts to the electron one
Eq.~(\ref{j_edge}) due to the $k^2$-dependence of the tensor
$\beta_{\rm BIA}$ for holes, Eq.~(\ref{beta_SIA_holes}).

\section{Results and Discussion}

Equations~(\ref{j_tot})~-~(\ref{xi}),~(\ref{beta_SIA_holes})~-~(\ref{gamma_SIA_holes}),
and~(\ref{j_BIA_holes})~-~(\ref{F_BIA_holes}) describe the
contributions to the circular photocurrent due to interband
optical transitions. It is seen that the symmetry of the system
determines the direction of the current. Indeed, according to
Eq.~(\ref{j_fin}), $j_i (\omega) \propto \beta_{li} \hat{o}_l$,
i.e. (i)~the current appears only under oblique light incidence in
(001)-grown QWs, and (ii)~for SIA the current ${\bm j}$ is
perpendicular to $\hat{{\bm o}}$, while for BIA the angle between
${\bm j}$ and $\hat{{\bm o}}$ is twice larger than the angle
between the axis [100] and $\hat{{\bm o}}$, see Fig.~\ref{f0}.

\begin{figure*} \epsfxsize=3in \epsfysize=4in
\centering{\epsfbox{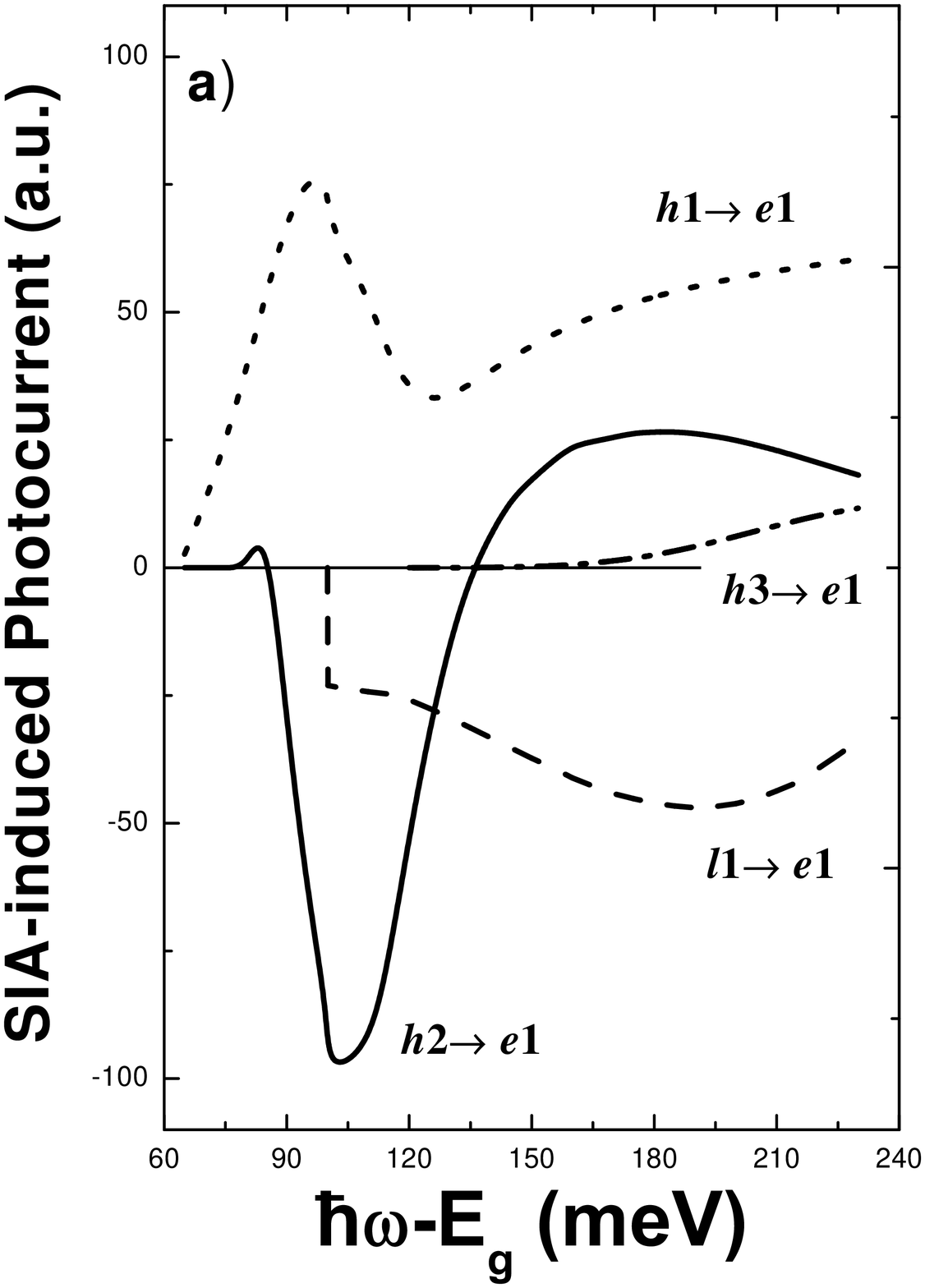}}   \epsfxsize=3in \epsfysize=4in
\centering{\epsfbox{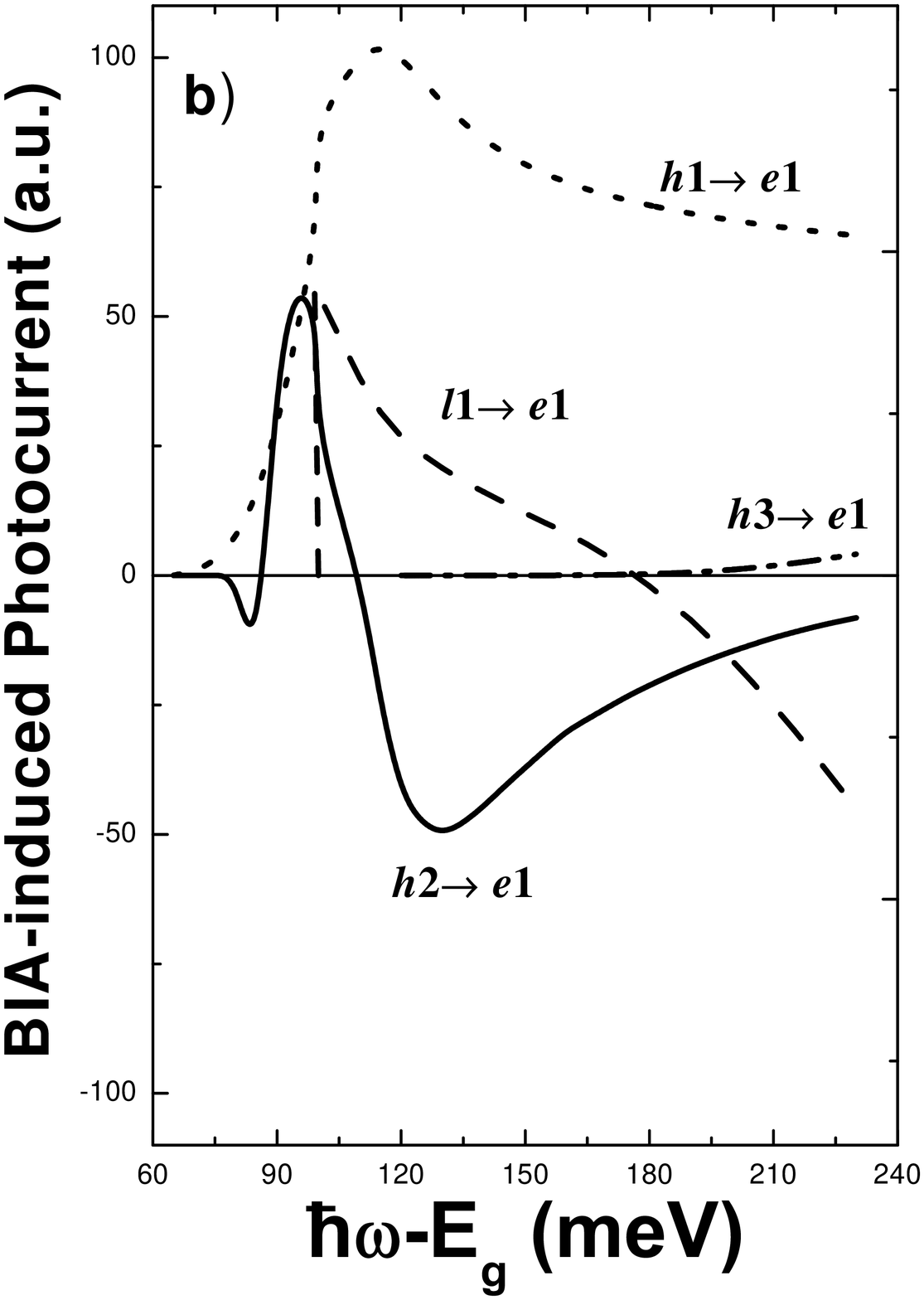}} \caption{\label{f2} Partial
contributions to the circular photocurrent for direct interband
transitions. Spin splitting of the electron states is due to SIA
(a) or BIA (b).}
\end{figure*}

\begin{figure}
\epsfxsize=3in \epsfysize=4in \centering{\epsfbox{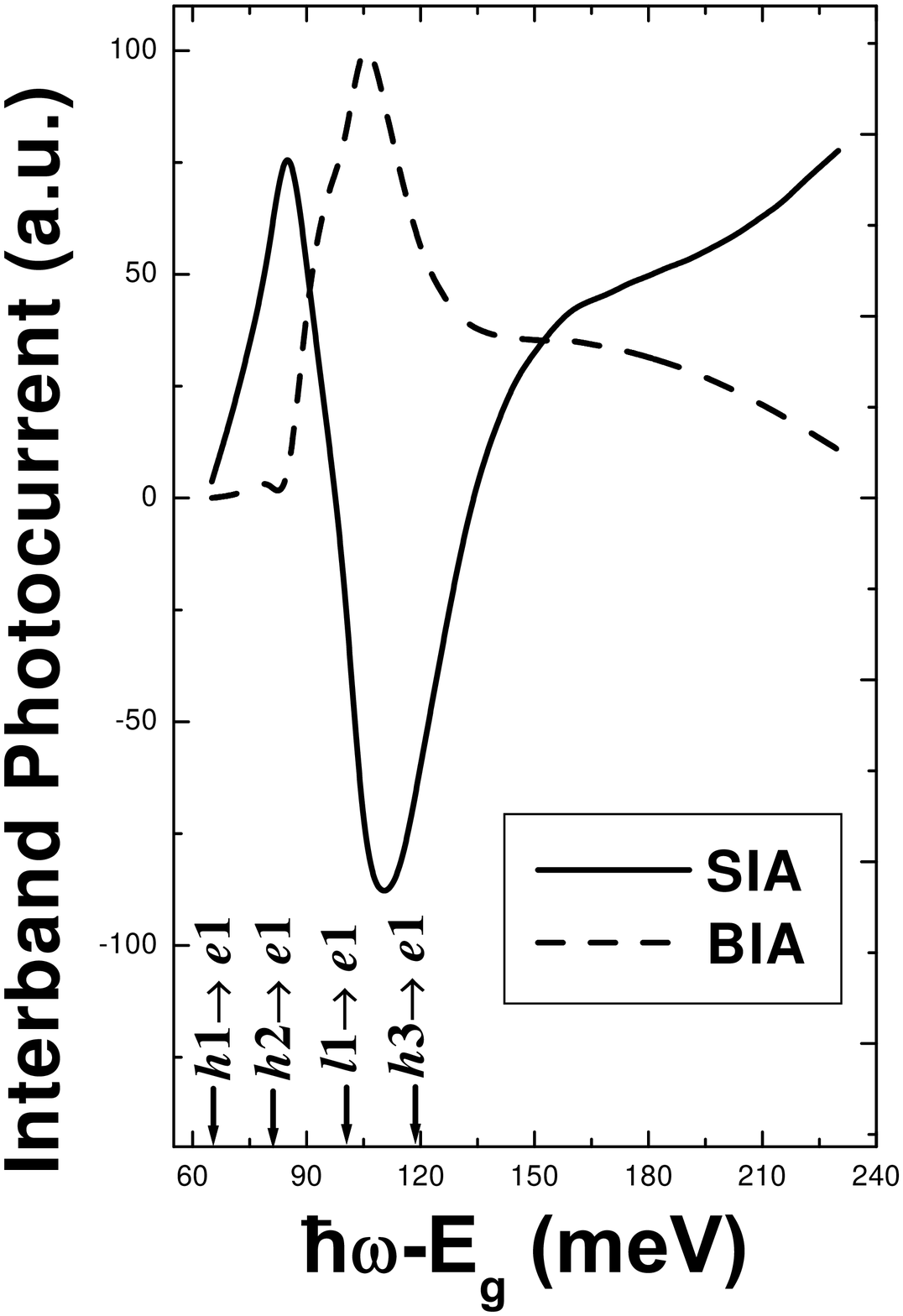}}
\caption{ \label{f1} Spectrum of the interband circular
photocurrent due to SIA (solid line) and BIA (dashed line)
electron spin splittings in a 100-\AA{} wide QW. The arrows
indicate the absorption edges for the four optical transitions.}
\end{figure}

The spectrum of the photocurrent is determined by the functions
$G(k_\omega)$. Figure~\ref{f2} presents the partial contributions
to this function for both  SIA and BIA electron spin splittings
calculated for a 100-\AA{} wide QW with infinitely-high barriers.
The interband transitions from the $h1$, $h2$, $l1$ and $h3$ hole
subbands to the ground electron subband $e1$ are taken into
account. The effective masses of the electron, heavy- and
light-holes are chosen to correspond to GaAs: $m_e=0.067 m_0$,
$m_{hh}=0.51 m_0$, $m_{lh}=0.082 m_0$. The momentum relaxation
times are assumed to be related by $\tau_h =2 \tau_e$ and
independent of the carrier energies.~\cite{tau}

The edge behavior of the photocurrent is due to the $h1 \to e1$
transitions. One can see the linear and quadratic raising of the
current near the absorption edge in accordance with
Eq.~(\ref{j_edge}). At higher energies, the spectra are determined
mainly by the $h1 \to e1$ and $h2 \to e1$ transitions. The
difference between BIA and SIA spectra is crucial: although in
both cases the current for the $h2 \to e1$ transitions is mainly
negative and has a minimum, in the BIA-induced photocurrent these
transitions give twice smaller contribution than $h1 \to e1$,
while for SIA-dominance they give the main contribution.

The total circular photocurrent caused by the four kinds of
optical transitions is presented in Fig.~\ref{f1}. The arrows in
Fig.~\ref{f1} indicate the points where the transitions start. One
can see that the BIA-induced circular photocurrent has a peak in
the spectrum, while in the SIA case the dip is present around the
same point. This photon energy corresponds to excitation of
carriers with $k \approx 2/a$, where $a$ is the quantum well
width. At this point the $h1$ and $h2$ energy dispersions have an
anticrossing. This results in a transformation of the hole wave
functions and, hence, substantial changes in the dependence
$G(k)$.

The main feature of Fig.~\ref{f1} is that the BIA-photocurrent has
no sign change in the given energy domain, while in the SIA-case
it has a sign-variable spectrum. This makes possible to determine
the structure symmetry by means of the photogalvanic measurements.

\section{Concluding Remarks}
The situations are possible when the both SIA and BIA are
present. The absolute value of the current in the case $\beta_{\rm
SIA} \cdot \beta_{\rm BIA} \neq 0$ is given by
\begin{equation}
\label{j_tot_modul} j (\omega) = \sqrt{j_{\rm BIA}^2(\omega) +
j_{\rm SIA}^2(\omega) \mp 2 j_{\rm BIA}(\omega) j_{\rm
SIA}(\omega) \sin{2 \phi}} \:,
\end{equation}
where $\phi$ is the angle between $\hat{{\bm o}}$ and the axis
[100]. The upper (lower) sign in Eq.~(\ref{j_tot_modul})
corresponds to $\beta_{\rm BIA} \cdot \beta_{\rm SIA} > 0$ ($<
0$). The direction of the photocurrent is given by the angle
$\psi$ between $[1\bar{1}0]$ and ${\bm j}$:
\begin{equation}
\label{psi} \tan{\psi} = {j_{\rm SIA}(\omega) + j_{\rm
BIA}(\omega) \over j_{\rm SIA}(\omega) - j_{\rm
BIA}(\omega)}\tan{(\phi - \pi/4)}\:.
\end{equation}
The  angular dependence~(\ref{j_tot_modul}) occurs due to
simultaneous presence of Rashba and Dresselhaus fields
($\beta_{\rm SIA} \cdot \beta_{\rm BIA} \neq 0$) by analogy with
the $\phi_{\bm k}$-dependence of the spin splitting. It should be
noted that $j_{\rm SIA}$ and $j_{\rm BIA}$ have different
excitation spectra that causes complicated $\omega$-dependences of
the total photocurrent absolute value and direction,
Eqs.~(\ref{j_tot_modul}),~(\ref{psi}).

The analysis shows that, under anisotropic scattering, the
``interference'' terms $\beta_{\rm BIA} \cdot \beta_{\rm SIA} /
(\beta_{\rm BIA}^2 + \beta_{\rm SIA}^2)$ appear in the total
photocurrent~Eq.~(\ref{j_tot}) even in the linear in $\beta$'s
regime. This is caused by coupling of the Fourier harmonics of the
velocity operator and density matrix in~Eq.~(\ref{j}).

In conclusion, we have developed a theory of the interband
circular photogalvanic effect in QWs. The cases when either SIA,
BIA or IIA dominates have been considered. It is shown that SIA
and BIA result in non-equal photocurrents in QWs. Their excitation
spectra have absolutely different shapes in the whole studied
range. This makes the interband circular photogalvanic effect a
unique highly sensitive tool for investigation of symmetry and
spin properties of QWs.

\section*{Acknowledgments}
The author thanks E.L.~Ivchenko for helpful discussions and
S.D.~Ganichev for a critical reading of the manuscript. This work
was financially supported by the RFBR, DFG, INTAS, ``Dynasty''
Foundation | ICFPM, and by the Programmes of RAS and Russian
Ministries of Science and Education.


\begin{thebibliography}{99}
\bibitem{zeta}Expressions for the real coefficients $N$, $W_\pm$, $Q_\pm$, and for the functions $\chi_\pm(z)$ are given in Ref.~\onlinecite{MPP} in the approximation of infinitely-high barriers.
\bibitem{tau} The contribution of the only ``allowed'' transition $l1 \to e1$ is proportional to ($\tau_h - \tau_e$) at $k=0$, therefore we take different values for $\tau_e$ and $\tau_h$ in order to obtain a step-like behavior of the spectrum. It is clearly seen in the corresponding curves in Figs.~\ref{f2}(a) and~\ref{f2}(b).
\end{thebibliography}
\end{document}